\documentclass[pra,reprint,doublecolumn,superscriptaddress, nobalancelastpage,dvipsnames]{revtex4-2}
\usepackage[dvipsnames]{xcolor}

\usepackage{tikz}
\usepackage{quantikz}
\usepackage{tikz-feynman}
\usetikzlibrary{shapes.misc, positioning, backgrounds}
\usepackage{pgfplots}
\usepackage[sort&compress]{natbib}
\usepackage{verbatim}
\usepackage{float}
\usepackage{sidecap}
\sidecaptionvpos{figure}{c}
\usepackage{lipsum}
\usepackage{placeins}
\usepackage{capt-of}
\usepackage{setspace}
\usepackage{dcolumn}
\usepackage{bm}
\usepackage{amssymb}
\usepackage{amsmath}
\usepackage{mathtools}
\usepackage{booktabs}
\usepackage{lipsum}  
\usepackage{physics}
\usepackage{multirow}
\usepackage[a4paper]{geometry}
\newgeometry{top=1.25cm,right=1cm,left=1.25cm,bottom=1.25cm}
\usepackage[colorlinks=true, urlcolor=blue, pdfborder={0 0 0}]{hyperref}
\hypersetup{linkcolor=blue,citecolor=BrickRed}
\DeclareMathOperator*{\argmin}{argmin}
\DeclareMathOperator*{\argmax}{argmax}
\newcommand{\x}{\boldsymbol{x}}

\DeclarePairedDelimiter{\ceil}{\lceil}{\rceil}

\definecolor{darkgreen}{rgb}{0.0,0.60,0.30} 
\definecolor{newred}{rgb}{1.00,0.70,0.70} 

\draft%
\begin{document}

\title{\LARGE{\normalfont{Drastic Circuit Depth Reductions with Preserved Adversarial Robustness \\ by Approximate Encoding for Quantum Machine Learning }}}

\author{Maxwell T. West} \email{westm2@student.unimelb.edu.au}  \affiliation{School of Physics, The University of Melbourne, Parkville, 3010, VIC, Australia}
\author{Azar C. Nakhl} \affiliation{School of Physics, The University of Melbourne, Parkville, 3010, VIC, Australia}
\author{Jamie Heredge} \affiliation{School of Physics, The University of Melbourne, Parkville, 3010, VIC, Australia}
\author{Floyd M. Creevey} \affiliation{School of Physics, The University of Melbourne, Parkville, 3010, VIC, Australia}
\author{Lloyd C.L. Hollenberg} \affiliation{School of Physics, The University of Melbourne, Parkville, 3010, VIC, Australia}
\author{Martin Sevior} \affiliation{School of Physics, The University of Melbourne, Parkville, 3010, VIC, Australia}
\author{Muhammad Usman} \email{musman@unimelb.edu.au} \affiliation{School of Physics, The University of Melbourne, Parkville, 3010, VIC, Australia}
\affiliation{Data61, CSIRO, Clayton, 3168, VIC, Australia}

\maketitle%
\onecolumngrid%

\noindent%
\textcolor{black}{\normalsize{\textbf{Quantum machine learning (QML) is emerging as an application of quantum computing with the potential to deliver quantum advantage, but its realisation for practical applications remains impeded by challenges. Amongst those, a key barrier is the computationally expensive task of encoding classical data into a quantum state, which could erase any prospective speed-ups over classical algorithms. In this work, we implement methods for the efficient preparation of quantum states representing encoded image data using variational, genetic and matrix product state based algorithms. Our results show that these methods can approximately prepare states to a level suitable for QML using circuits two orders of magnitude shallower than a standard state preparation implementation, obtaining drastic savings in circuit depth and gate count without unduly sacrificing classification accuracy. Additionally, the QML models trained and evaluated on approximately encoded data  display an increased robustness to adversarially generated input data perturbations. This partial alleviation of adversarial vulnerability, possible due to the ``drowning out'' of adversarial perturbations while retaining the meaningful large-scale features of the data, constitutes a considerable benefit for approximate state preparation in addition to lessening the requirements of the quantum hardware. Our results, based on simulations and experiments on IBM quantum devices, highlight a promising pathway for the future implementation of accurate and robust QML models on complex datasets relevant for practical applications, bringing the possibility of NISQ-era QML advantage closer to reality.}}}
\\ \\ \\
\twocolumngrid%

\begin{figure*}
\includegraphics[width=0.9\textwidth]{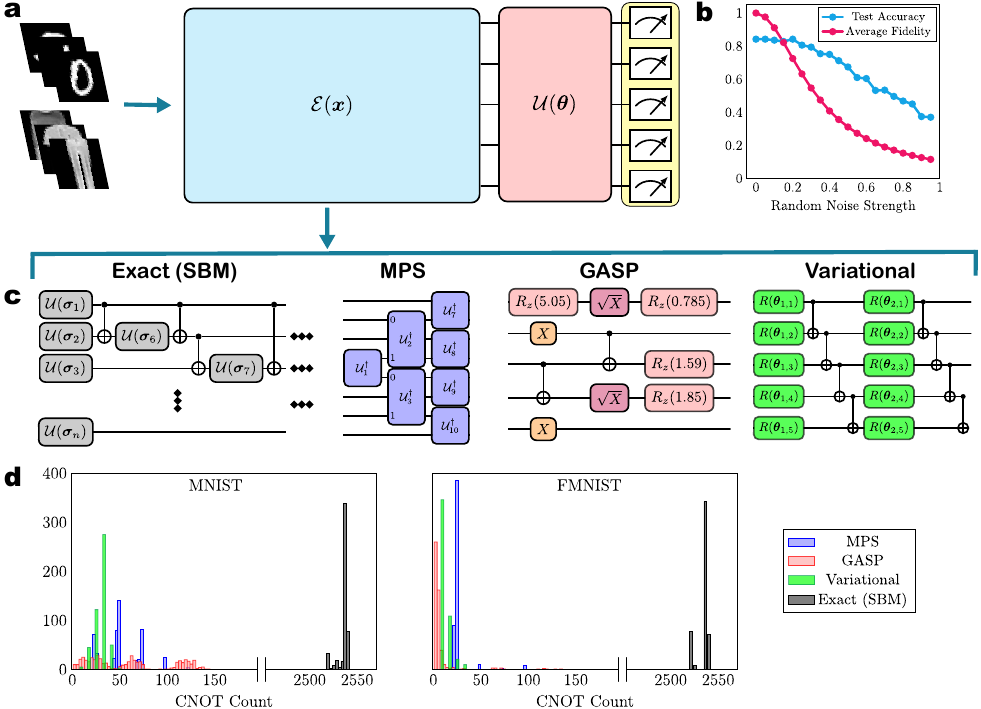}
  \caption{\textbf{Approximate State Preparation Techniques.} \textbf{a.} A schematic diagram of a typical QML framework. Generic $n$-qubit quantum state preparation ($\mathcal{E}(\x)$) has an $\mathcal{O}\left(2^n\right)$ runtime, 
  while the depth of variational circuits ($\mathcal{U}({\theta})$) are commonly restricted to $\mathcal{O}\left(\mathrm{polylog}\ n\right)$ to avoid barren plateaus~\cite{mcclean2018barren,holmes2022connecting,cerezo2021cost,pesah2021absence}. Without mitigation, therefore, for large $n$ the initial state preparation stage will dominate the runtime of the algorithm and eliminate any potential quantum advantage afforded by the variational model. \textbf{b.} The accuracy of a quantum variational classifier  on 500 test examples from the MNIST dataset is plotted in the presence of uniformly random perturbations to the image pixels, as well as the average fidelity of the resulting encoded states (see Equation~\ref{eq:amp_enc}) with respect to the clean images. We observe significant robustness to these perturbations, with the accuracy remaining roughly constant even as the fidelity of the noisy states drops to 60\%. \textbf{c.} Typical circuits resulting from the SBM technique employed in the Qiskit state preparation routine as well as the MPS, GASP and variational methods. While the SBM, MPS and variational approaches produce circuits with a deterministic structure, GASP  can find very different looking circuits for different target states, occasionally containing only a few CNOT gates. \textbf{d.} Histogram distributions of the number of CNOT gates required to construct 500 9-qubit states from the MNIST and FMNIST datasets by the methods considered here. The approximate methods are given a target fidelity of 60\%, and require two orders of magnitude fewer gates than the method that Qiskit~\cite{shende2005synthesis,qiskit} employs to construct the state exactly.
  }
\label{fig:1}
\end{figure*}

\noindent
\large{\textbf{Introduction}}
\normalsize
\\ \\
\noindent
The incredible capabilities of Transformer based models~\cite{bubeck2023sparks,copilot,rombach2022high,radford2022robust,ho2022imagen} has provoked society-wide interest in artificial intelligence (AI) and machine learning (ML), which is increasingly moving beyond academic and scientific 
applications and into business, industrial and military use cases.
Concurrently, the emergence of programmable quantum computers has led to intense interest in the prospect
of quantum machine learning (QML)~\cite{biamonte2017quantum,beer2020training, havlivcek2019supervised, schuld2019quantum,qcnn,tsang2022hybrid,schuld2021supervised,westboosted2023,schatzki2022theoretical,liu2021rigorous,huang2022quantum, west2022reflection} -- the study of ML algorithms which exploit the capabilities of quantum computers.
Given the rapid proliferation of ML technology, any speedups, enhancements to robustness or other advantages which can be afforded by quantum computing have the potential to be highly impactful. Indeed QML models have been shown to in principle possess the ability to make use of classically intractable features of data to outperform conventional classical methods through exponential speed-ups~\cite{liu2021rigorous} and enhanced resilience to adversarial attacks~\cite{west2023benchmarking}. However, it is unclear whether such features will generally prove useful for generic classification tasks, particularly classical data 
which has no inherently quantum mechanical source or structure. 
Before it can be processed by a quantum computer, such data must first be encoded into a quantum state, 
a generically $\mathcal{O}(2^{n_{\mathrm{qubits}}})$ process~\cite{shende2005synthesis} which has the potential to dominate the runtime of a QML algorithm and negate any potential quantum advantage (see Figure~\ref{fig:1}(a)). Quantum state preparation~\cite{niemann2016logic,shende2005synthesis,10.5555/1131481.1131569,daskin2011decomposition,creevey2023gasp} is therefore often the first and most computationally expensive subroutine of a QML model, but remains a comparatively understudied component of the QML pipeline. With the quantum devices of the current generation offering only limited capabilities in terms of both number of qubits and gate fidelities, the implementation and benchmarking of efficient quantum state preparation techniques 
(reducing circuit depths) 
such as performed here is an important step towards improving the prospects of QML models for datasets of practical interest.

\begin{figure*}
\includegraphics[width=0.95\textwidth]{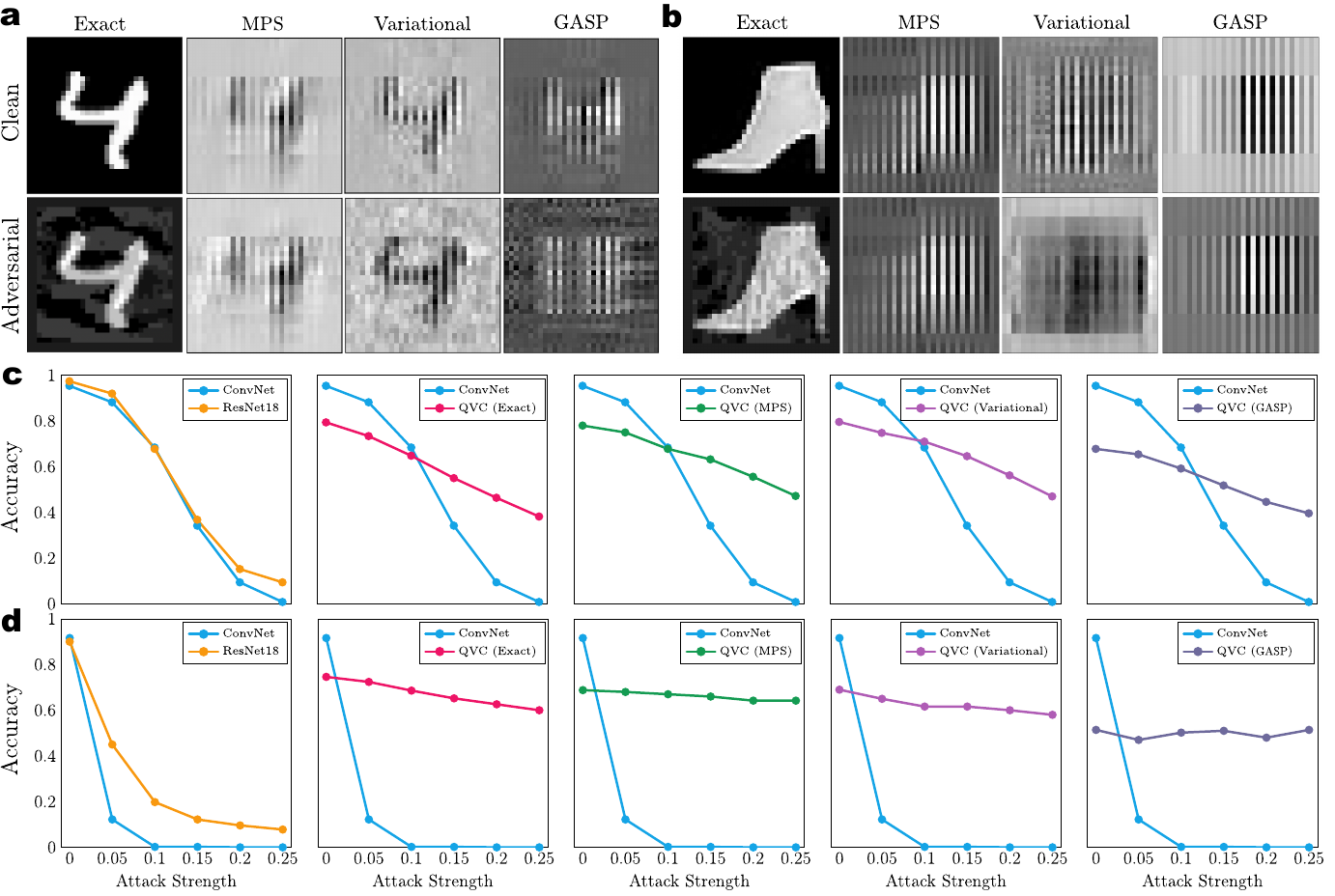}
\caption{\textbf{Adversarially Robust Quantum Machine Learning.} \textbf{a}. Given an approximately prepared state, we can use Equation~\ref{eq:amp_enc} to reconstruct the image it represents. Here this is done for an example image from the MNIST dataset (top left) and a corresponding  adversarial example  (bottom left). Despite the clear loss of visual quality in the approximate images, the QVCs retain their level of performance to a remarkable degree. This is especially true in the adversarial case; adversarial perturbations are carefully constructed, and a low fidelity approximation can overlook the fine details which are required to trick the network~\cite{li2019certified}. \textbf{b}. The case of the FMNIST dataset is similar. Further examples are shown in Supplementary Figures~\ref{fig:noisy_ims}--\ref{fig:alt_encoding}, where slight variations on the encoding scheme are discussed. \textbf{c}. The accuracies obtained by a classical convolutional neural network (CNN), ResNet18~\cite{he2016deep} and the QVCs on adversarial images from the MNIST dataset generated by attacking the CNN. As reported in Ref.~\cite{west2023benchmarking} we find that the QVCs display considerable robustness to the classically generated adversarial images. At an attack strength of zero (i.e. unperturbed images) we find that the approximate state preparation techniques can be used to obtain comparable accuracy to a model using the exact states. The MPS and variational approaches more readily obtain higher fidelities than  GASP, accounting for their improved performance (see Supplementary Figures~\ref{fig:training_val_accs}--\ref{fig:val_fidelities}). \textbf{d.} Similarly, we  train the models on the FMNIST dataset and construct adversarial examples by attacking the CNN. We again find significantly improved robustness on the part of the QVCs. } 
\label{fig:adv_ims}
\end{figure*}

\noindent
This work demonstrates that one can preserve the accuracy and increase the adversarial robustness of QML models classifying image data by moving to approximate  data encoding schemes, while simultaneously reducing the encoding circuit complexities by orders of magnitude, providing a crucial advantage in experimentally implementing QML on noisy quantum hardware platforms. 
The ability to do this stems from the generic resilience of machine learning models to random perturbations of their inputs~\cite{szegedy2013intriguing}. For example, Figure~\ref{fig:1}(b) shows that our QML models are capable of
maintaining their accuracy on noisy image data (with fidelities as low as 60\%) encoded exactly into quantum states. This indicates that an approximation to the input quantum state compromising fidelity but resulting in an easier circuit preparation with reduced depths (speeding up the state preparation) may have only a minor impact on the QML accuracy. Motivated by this remarkable property, we consider three independent approximate state preparation methods (see  Figure~\ref{fig:1}(c)), based on matrix product states (MPS)~\cite{ran_encoding_2020,cramer_efficient_2010,schon_sequential_2005}, genetic algorithms (specifically, the GASP method of Ref.~\cite{creevey2023gasp}) and variational circuits. Our results show that all three methods are capable of preparing states representing images from the standard MNIST~\cite{mnist} and FMNIST~\cite{xiao2017fashion} datasets using circuits two orders of magnitude shallower than the high fidelity states constructed by the exact algorithm (SBM) employed by Qiskit~\cite{shende2005synthesis,qiskit} as shown in Figure~\ref{fig:1}(d), while largely maintaining the accuracies achieved by the exact method. Furthermore, on a simple dataset of $8\cp 8$ images, the savings in circuit depth afforded by the approximations are sufficient to perform classification on the device \textit{ibm\_algiers} with up to 96\% accuracy, whereas the exact encoding method leads to circuits dominated by noise and performance indistinguishable from random guessing. \\ \\
\noindent
The adversarial vulnerability of QML models has recently become a topic of significant research interest due to the serious concerns resulting from the increasing incorporation of machine learning in security sensitive applications~\cite{west2023towards}. It has been suggested empirically~\cite{west2023benchmarking} that QML models may provide enhanced resilience to (classical) transferred adversarial attacks, providing an advantage to early adopters of quantum computers. Appealingly, this argument is independent of the whether the classically intractable features which will generally be learnt by quantum networks are superior for the classification task at hand, a condition which is by no means guaranteed~\cite{west2023benchmarking}. We find that the adversarial robustness of the approximate models considered in this work are actually slightly increased over their exact counterparts, due to the errors in the state preparation process ``drowning out'' the carefully constructed adversarial perturbations (see Figure~\ref{fig:adv_ims}(a)). This is reminiscent of well-known results in classical machine learning~\cite{cohen2019certified,li2019certified,lecuyer2019certified}, which involve intentionally introducing random noise in order to combat adversarial perturbations. Here, however, we obtain a similar benefit through the pseudo-randomness introduced by the approximate state preparation, in addition to the primary motivation, reducing  circuit depths. These results, in which drastic improvements to the circuit depth of the state preparation component of the models in fact leads to a slight increase in the resilience to adversarial attacks, while almost preserving classification accuracy, highlight a promising pathway for the future development and deployment of QML models in practical applications where security and robustness will be key parameters of interest. 
\\ \\
\noindent
\large{\textbf{Approximate State Preparation}}
\normalsize
\\ \\
\noindent
Our investigation focuses on the problem of image classification, with images drawn from the MNIST~\cite{mnist} and FMNIST~\cite{xiao2017fashion} datasets (as well as, later, a simple dataset constructed for the purpose of running small QML models on real devices). The first step of our QML models is an encoding unitary $\mathcal{E}(\x)$ which, given an image $\x$ (represented as a vector of pixel values) creates a quantum state $\ket{\psi(\x)}=\mathcal{E}(\x)\ket{0}^{\otimes n}$. Due to the high dimensionality of the data (images from MNIST and FMNIST contain 28$\times$28 pixels and so $\x\in\mathbb{R}^{784}$), the current limitations of quantum hardware and the difficulty of simulating quantum computers possessing large numbers of qubits, we employ the method of amplitude encoding, which is highly qubit efficient compared to other common encoding techniques such as angle encoding~\cite{larose2020robust}. Moreover, as the pixel values are real numbers, we can encode two of them into each of the complex amplitudes of the state $\ket{\psi(\x)}$.
Our mapping procedure is therefore
\begin{equation}
 \x \mapsto \ket{\psi(\x)}=\frac{1}{\norm{\x}^2}\sum_{j=0}^{2^n-1} \left(x_{2j}+ix_{2j+1}\right)\ket{j}   \label{eq:amp_enc}
\end{equation}
and requires $n=\ceil{\log_2\left(28\times 28\right)}-1=9$ qubits.
We note that due to the global phase ambiguity and the normalisation constraint of a quantum state this method of encoding cannot distinguish between two images that differ only by a constant factor, however for the image classification tasks considered here this does not prove to be an important deficiency. Although this method of encoding is very efficient in terms of the required number of qubits, it involves the preparation of highly entangled states, the circuit depth required for which is generically exponential in the number of qubits. For the $n=9$ states we consider here, the resulting circuit depths of $\sim$$2^9=512$ put the preparation of $\ket{\psi(\x)}$ with high fidelity beyond the capability of the noisy quantum hardware available today. \\
\\
\noindent
Interestingly, the QML models we employ (see Figure~\ref{fig:1}(a)) display considerable robustness to random perturbations of the input states $\ket{\psi(\x)}$, which suggests that preparing them with high fidelity is unnecessary. In Figure~\ref{fig:1}(b) we add uniformly random (not to be confused with adversarial) noise to a set of test images from the MNIST dataset before encoding, and plot both the accuracy of our model on the resulting images, and the average fidelity of the encoded states of the noisy images with respect to the encoded states of the clean images, as a function of the the level of random noise applied. We find that the accuracy of the classifier remains steady even as the fidelities of the noisy states drops to $\sim$$60\%-70\%$. This is an extremely encouraging result, as the preparation of a quantum state to this level of fidelity is a drastically easier task than preparation to high fidelity. 
\\ \\
\noindent
Motivated by the notion that accurate QML models can be implemented on very noisy encoded states we implement three methods of approximate state preparation and benchmark them by both their raw accuracy and adversarial robustness: a deterministic technique for constructing a matrix product state (MPS) approximation to the target state~\cite{cramer_efficient_2010}, and two machine learning based methods which attempt to learn unitaries that perform the state preparation, relying in one case on genetic algorithms (GASP)~\cite{creevey2023gasp} and in the other on variational quantum circuits. Full details of each of these approaches are discussed below. When applied to the 9 qubit states representing images from the MNIST and FMNIST datasets these methods lead to encoding circuits with two orders of magnitude fewer entangling gates (Figure~\ref{fig:1} (d)) when compared to the (exact) SBM algorithm implemented in Qiskit~\cite{shende2005synthesis,qiskit}.
\\ \\
\noindent
\textbf{Matrix Product State Encoding:} Matrix Product States (MPS) are a tensor network representation of quantum systems that
have found significant application in the simulation of quantum algorithms as a result of their ability to represent and manipulate states of relatively low bipartite entanglement in a memory efficient manner~\cite{vidal_efficient_2003}. States represented in MPS form may be decomposed in such a way that they are disentangled entirely using sequential $k$-local operators~\cite{ran_encoding_2020,schon_sequential_2005}, where $k=\lceil\log_2(D)\rceil+1$ is a function of the maximum internal bond dimension of the state $D$. This procedure has applications in quantum state tomography~\cite{cramer_efficient_2010} and image classification~\cite{dilip2022data}, and when run in reverse  reconstructs the original state using only $k$-local gates. \\

\noindent

\noindent
The procedure begins by determining the eigen-decomposition of the reduced density matrix over the first $k$ sites,
\begin{equation}    
    \rho_{(0,1,\dots,k)} = \sum_{i=1}^{2^k} \lambda_i \dyad{\phi_i} \label{eq:red}
\end{equation}
Reduced density matrices can be determined efficiently given an MPS in canonical form~\cite{schollwock_density-matrix_2011}. Moreover, the rank of the reduced density matrix is bounded by $2^{k-1}$~\cite{cramer_efficient_2010}, and consequently $\lambda_i = 0$ for all $i > 2^{k-1}$. Given this, following from Equation \ref{eq:red} one can construct the following unitary operator,
\begin{equation}    
    U_{0} = \sum_{i=1}^{2^k} \dyad{i}{\phi_i} \label{eq:unit}
\end{equation}
where the eigenvectors $\ket{\phi_i}$ of $\rho_{(0,1,\dots,k)}$ are ordered in decreasing order of eigenvalue. Acting with $U_0$ on the system gives
\begin{equation}
    U_0 \rho_{(0,1,\dots,k)} U_0^\dagger = \sum_{i=1}^{2^{k-1}} \lambda_i \dyad{i} \label{eq:opu}
\end{equation} 
Hence we find that the operator $U_0$ completely disentangles the first qubit from the rest of the system. This procedure is then repeated for subsequent qubits until the entirety of the system becomes a product state. \\ \\
This procedure will successfully prepare a set of unitary operators which create the desired state, however the size of the unitary operators will be in general exponentially large, as dictated by the maximum bond dimension $D$ of the MPS. If we restrict ourselves to only considering the case  $k=2$ irrespective of the entanglement of the state, then in general the reduced density matrices will be of full rank and hence Equation \ref{eq:opu} will no longer hold as there will be $2^k$ non-zero eigenvalues. However it can be demonstrated that the application of the operator in Equation \ref{eq:unit} will in general result in an increase in the magnitude of the amplitude of the $\ket{0}^{\otimes N}$ state~\cite{nakhl2021simulating}. Given this one may define a heuristic procedure where Equations \ref{eq:red}-\ref{eq:opu} are performed across the system multiple times with a fixed $k$ until some threshold fidelity is met. 
\\ \\
\noindent
Another consequence of this approximate disentangling procedure is that one can relax the requirement that each subsystem is processed in sequence and instead several subsystems may be considered at once, with the only requirement being that there must be an overlap between all the subsystems by the end of the iteration. As such we propose a circuit structure where all pairs of adjacent qubits are processed in parallel, with subsequent operations which join each of the disparate subsets (see Figure \ref{fig:1}(c)). We use $k=2$ as the size of our unitary operators for the entirety of this work as they may be decomposed with at most 3 CNOT gates~\cite{iten_quantum_2016}. The individual 2-qubit unitary operators are decomposed using IBM's Qiskit SDK~\cite{qiskit}.
\\ \\
\noindent
\textbf{Genetic Algorithm for State Preparation (GASP):} Genetic algorithms are a classical optimisation technique that aim to mimic the process of natural selection in nature \cite{whitley_genetic_1994} and have been used with varying degrees of success in quantum state preparation on quantum computers \cite{spector_genetic_1998, creevey2023gasp, rindell_generating_2022}. The main advantage of genetic algorithms for state preparation is their ability to reduce the number of CNOT gates required in a state preparation circuit, as well as reduce the total gate count. This results in circuits that are more noise-resistant, an important feature in the current NISQ era of quantum computing. The method used in this paper is the Genetic Algorithm for State Preparation (GASP), introduced in Ref.~\cite{creevey2023gasp}.
\\ \\
\noindent
The method begins by defining a target state vector $\ket{\psi_{\rm{target}}}$ to be evolved, in this case a state representing an encoded image as in Equation \ref{eq:amp_enc}. For this given target state vector a population of individuals, $|\psi(\vec{\theta})_{P_i}\rangle$, is produced, with a certain number of `genes', which represent quantum gates, drawn from a specified gate set, in this case, \{$X, \sqrt{X}, R_x(\theta), R_y(\theta), R_z(\theta), \mathrm{CNOT}\}$, where CNOT gates can only be placed between nearest neighbour qubits. The fitness of each individual in this population is then assessed with a defined fitness function, in this case, $|\langle\psi_{\rm{target}}|\psi(\vec{\theta})_{P_i}\rangle|^2$. The genetic operators of crossover and mutation are then applied to the population. Crossover in this case is a $1$-point crossover, taking half the genetic information (quantum gates) from one parent, and half from the other. Mutation corresponds to changing each gene (gate) in an individual with a probability $p = 5\%$. SLSQP optimisation~\cite{kraft1988software} is then run on each individual to allow the highest fidelity for each individual (circuit structure) to be achieved. The genetic operator of selection is then applied to the population, which in this case is roulette wheel selection, in which each individual is assigned a probability of selection based on their fidelity, with the fittest individuals having the highest probability of selection. This method allows for low-fitness, but `lucky', individuals to be selected, thus maintaining a high degree of genetic diversity. This iteratively increases the fitness of the population. These steps are repeated until the desired target fitness is achieved, or there have been 200 generations without the desired fitness achieved. If the desired target fitness is achieved, the number of gates in the circuit is reduced, so as to achieve the shortest circuit required to produce the desired target state vector. If the desired fitness is not achieved, the number of genes in the individuals is increased, allowing more complex states to be evolved. This increase and decrease of genes is implemented in a binary search (starting from an initial population with $N_{\mathrm{genes}}=64$) to exponentially reduce the complexity of finding the optimal number of genes compared to a linear search starting from $N_{\mathrm{genes}}=1$. 
\\ \\
\noindent
Unlike the other methods considered in this work, the circuits discovered by the GASP algorithm are not restricted to be of a repeating, deterministic form (see Figure~\ref{fig:1}(c)). This allows the algorithm to occasionally find circuits consisting of only a few CNOT gates, even for complicated $9$-qubit states (see Figure~\ref{fig:1}(d)), by exploiting the specific details of individual statevectors.
\\ \\
\noindent
\textbf{Variational Encoding:} Our final method consists of a variational circuit $\mathcal{V}(\boldsymbol{\theta})$ which is optimised to maximise the fidelity $\abs{\bra{\psi_{\mathrm{targ}}}\mathcal{V}(\boldsymbol{\theta})\ket{0}}^2$ of the resulting state with the target state $\ket{\psi_{\mathrm{targ}}}$. The variational circuit is composed of a sequence of layers, with each layer consisting of an arbitrary rotation on each qubit $i$, followed by a chain of CNOT gates applied between adjacent qubits (see Figure~\ref{fig:1}(f)). Initially $\mathcal{V}(\boldsymbol{\theta})$ is  composed of a single such layer, and its parameters are optimised with 100 steps of the Adam optimiser~\cite{adam}. If this fails to produce a state that reaches the given target fidelity then an additional layer is added and the optimisation repeated (up to a maximum of 25 layers). Despite the simplicity of this method we find that it is able to prepare the target states using resources commensurate with our other two approaches, generally requiring only a few variational layers to achieve 60\% fidelity (see Figure~\ref{fig:1}(c)).
\\ \\
\noindent
\large{\textbf{Adversarial Quantum Machine Learning}}
\normalsize
\\ \\
\noindent
Despite the increasingly impressive successes of machine learning across a large number of domains, the prospect of automating high-impact tasks with a low tolerance for failure has been hindered by the lack of explainability of ML models, which typically produce their outputs via computational processes that are inscrutable to humans~\cite{zhang2021survey}. Among other difficulties, the complicated and opaque nature of the decision-making process of these models leaves them open to surprising failure modes~\cite{szegedy2013intriguing}. Indeed, an unexpected discovery of recent years has been the extreme susceptibility of artificial neural networks to careful tampering with their inputs (an ``adversarial attack''), which can cause dramatic failures even in  sophisticated high-performance 
models~\cite{biggio2013evasion,szegedy2013intriguing,10.1145/2046684.2046692,kurakin2016adversarial,goldwasser2022planting,wong2020fast, pgd, goodfellow2018making, miller2020adversarial, cohen2019certified, bai2021recent}.
Moreover, adversarial attacks tend to \textit{transfer} between neural networks -- an adversarial example constructed to deceive a specific
network can also fool other, entirely independent networks~\cite{szegedy2013intriguing,ilyas2019adversarial,fgsm,tsipras2018robustness}.
This significantly increases the threat posed by adversarial attacks in practice, as it means that one can trick an external model without being privy to the precise details of its architecture. As part of the ongoing effort to determine the potential benefits and practicality of QML, the adversarial robustness of quantum 
classifiers has therefore become the subject of much recent 
attention~\cite{west2023towards,lu2020quantum,quantum_cm,du2021quantum,guan2021robustness,weber2021optimal,ren2022experimental,liao2021robust,kehoe2021defence,west2023benchmarking}.
The evidence compiled to date indicates that QML models will also suffer from the existence of adversarial examples, 
as a result of counter-intuitive geometrical properties of the Hilbert spaces in which their classifications takes place~\cite{quantum_cm}.
Recent work~\cite{west2023benchmarking}, however, has suggested that quantum classifiers may demonstrate significant 
resilience to adversarial examples generated by attacking a classical network and then transferring the results to the quantum network.
This is due to the QML models learning a different set of features to those generically learnt by the classical networks,
and is potentially a new avenue through which to demonstrate quantum advantage in machine learning: quantum networks 
which offer enhanced robustness to classical adversarial attacks~\cite{west2023benchmarking}. 
\\ \\
\noindent
Our QML models follow a standard~\cite{west2023benchmarking} three-step paradigm: a data dependent encoding unitary $\mathcal{E}(\x)$ (implemented variously by our MPS, variational and GASP strategies), followed by a trainable variational circuit
$\mathcal{U}({\theta})$, followed by the measurement of a set of observables $\{M_j\}_{j=1}^{n_{\mathrm{classes}}}$.
Having discussed the initial data encoding stage in detail in the previous section, we now turn to the design of the remaining aspects of our models, and their susceptibility to adversarially manipulated data.
\\ \\
\noindent
The variational component of our models consists of a sequence of repeated layers. Each layer comprises (trainable) arbitrary rotations on each qubit, followed by a chain of CNOT gates between neighbouring qubits.
The prediction $\hat{y}_{\theta}(\x)$ of a model on a data point $\x$ is defined to be the class corresponding to the measured
observable with the highest expectation value, i.e.,
\begin{equation}
  \hat{y}_{\theta}(\x) = \argmax_j \bra{0}^{\otimes n}\mathcal{E}(\x)^\dagger\mathcal{U}({\theta})^\dagger M_j \mathcal{U}(\theta)\mathcal{E}(\x)\ket{0}^{\otimes n}\label{eq:prediction}
\end{equation}
We can subsequently assign a probability $p(j|\x)$ that the input $\x$ belongs to the
class with index $j$ via a standard softmax normalisation:
\begin{equation}
  p_\theta(j|\x)  = \frac{\exp\left(\bra{\psi(\x)}\mathcal{U}({\theta})^\dagger M_j \mathcal{U}(\theta)\ket{\psi(\x)}\right)}{\sum_k \exp\left(\bra{\psi(\x)}\mathcal{U}({\theta})^\dagger M_k \mathcal{U}(\theta)\ket{\psi(\x)}\right)}\label{eq:probs}
\end{equation}
where we have written $\ket{\psi(\x)}=\mathcal{E}(\x)\ket{0}^{\otimes n}$ to denote the encoded state corresponding to the data point  $\x$. Throughout this work we take the $j$th measurement observable $M_j$ to be the Pauli $z$ operator acting on the $j$th qubit. 
During training, the variational parameters  $\theta$ are updated to minimise the empirical cross-entropy loss~\cite{zhang2018generalized} on the training set,
\begin{equation}
\label{eq:loss_min}
  \mathcal{L}_{N_{\mathrm{train}}} (\boldsymbol {\theta}) = \frac{1}{N_{\mathrm{train}}} \sum _ {  i = 1  }  ^{  N_{\mathrm{train}} } -\log \left(p_\theta\left(y_i|\x_i\right) \right)
\end{equation}
where $N_{\mathrm{train}}$ is the number of training examples and $y_i$ is the label of the datapoint $\x_i$. We train the models by seeking parameters
   $\boldsymbol {\theta}^* = \argmin_{\boldsymbol {\theta}} \mathcal{L}_{N_{\mathrm{train}}} (\boldsymbol {\theta})$
which minimise the empirical loss, where we carry out the optimisation using the Adam optimiser~\cite{adam} with a learning rate of 1e-3 and a training set of 30,000 images. For each state preparation technique a model is trained and tested on data prepared using that method, with the exception of GASP, which is tested on models trained using variationally prepared states due to limited computational resources.
The accuracies obtained on the validation set throughout the training process for each of the considered state preparation methods are shown in Supplementary Figure~\ref{fig:training_val_accs}.

\begin{figure*}[t]
\includegraphics[width=\textwidth]{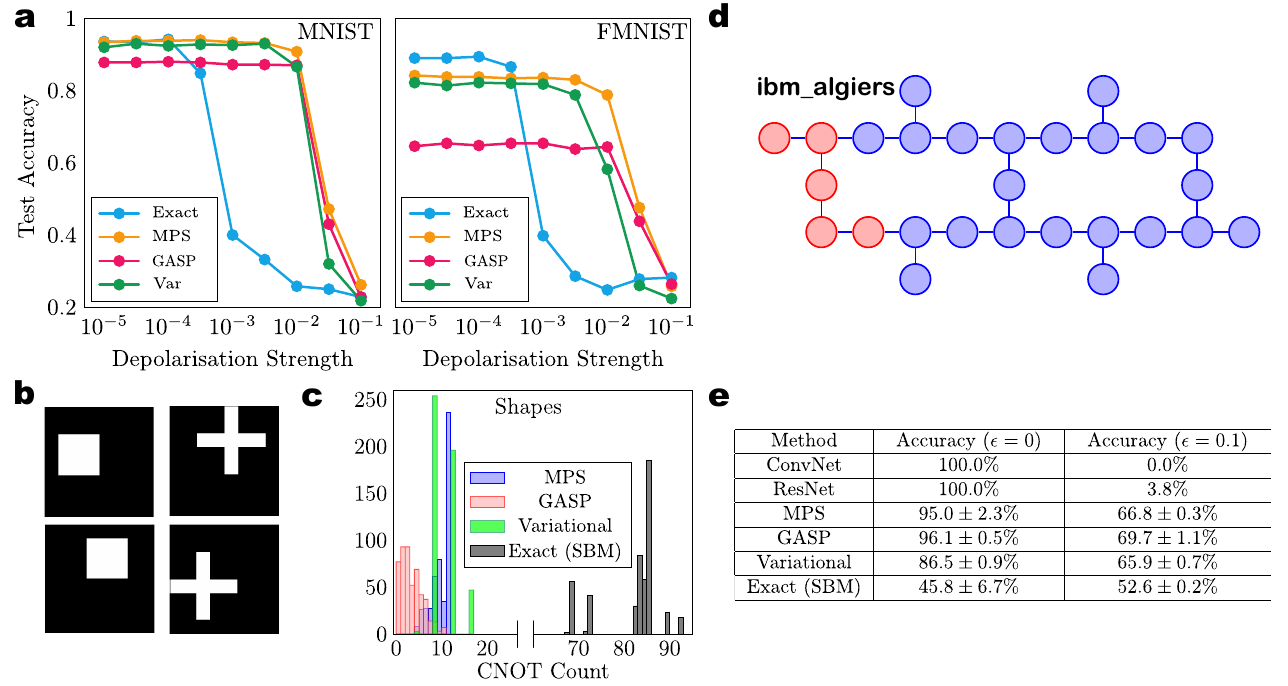}
\caption{\textbf{Experimental Demonstration.} 
\textbf{a.} 
The accuracies of standard 20-layer QVCs  in the presence of depolarisation noise acting on each of the two-qubit gates in the circuit. Due to their shorter depths, the QVCs employing approximate state encoding methods are able to maintain their accuracies up to a more realistic noise level of $p_{\mathrm{depol}}\sim 10^{-2}$. In this example the QVCs have been trained to classify the first four classes of the MNIST and FMNIST datasets.
\textbf{b.} 
We introduce a simple, two class dataset ``Shapes'' consisting of 8$\cp$8 images. Using our encoding scheme (Equation~\ref{eq:amp_enc}) this allows us to encode an image into a five qubit state. Images from this dataset consist of ``squares'' and ``crosses'' positioned randomly on an 8$\cp$8 grid.
\textbf{c.} A histogram of the number of CNOTs needed by the various methods to construct a set of 500 test images from the Shapes dataset to a target of 70\% fidelity. As in the cases of MNIST and FMNIST (see Figure~\ref{fig:1}(c)), we find that our three approximate methods considerably outperform the default Qiskit behaviour. 
\textbf{d.}
We evaluate our QML models on this dataset on real hardware, the 27 qubit superconducting system \textit{ibm\_algiers}. The topology of \textit{ibm\_algiers} is shown, with the five qubits employed highlighted in red. 
 \textbf{e.} 
 The accuracy on 500 test examples from the Shapes dataset for different state preparation methods, evaluated on \textit{ibm\_algiers} in the presence of adversarial attacks on ConvNet. We run each experiments three times and report the mean and standard deviation. While the default Qiskit state preparation routine produces deep circuits which are overwhelmed by noise,  the savings in circuit depth realised by our approximate methods are sufficient to allow our QVC to classify the data with high accuracy. All QVCs again display considerable robustness to adversarial attacks generated by attacking a convolutional neural network and transferring the resulting images, but this time on a real device in the presence of complicated hardware noise.}
\label{fig:shapes}
\end{figure*}

\noindent
The process of generating an adversarial example with respect to a given ML model (``an adversarial attack''),  
on the other hand, consists of an attempt to \textit{maximise} the loss function~\cite{pgd,fgsm,auto}. 
We consider the model parameters $\boldsymbol {\theta}$  to be fixed at their trained values $\boldsymbol {\theta}^*$,  and,  given a 
target example $  \boldsymbol { x}  $ with label $y$  and a region $ \Delta  $ consisting of ``small'' inputs  
(with respect to a relevant metric, see below), search for a perturbation 
\begin{equation}
  \boldsymbol{\delta}_{\mathrm{adv}}  = \argmin_{ \boldsymbol{\delta}  \in  \Delta}\ p_{\theta^*} (y|\x+\boldsymbol{\delta}) \label{eq:delta_adv}
\end{equation}
where  $ y $  is  the label associated with $  \boldsymbol { x}  $. That is, we look for a minor modification we can make to the input $\x$ which minimises the probability that the model assigns the label $y$ to the new input. If the perturbation is  small, then, by assumption, the true label of the modified datapoint is also $y$, and so if the procedure is successful the model has been tricked into making a misclassification. Typically, one takes $  \Delta  $ to be the space of inputs with $l_p$ 
norm $\norm{\boldsymbol{\delta}}_p= \left(  \sum _ { i = 1 } ^{  n} \abs{ \delta  _ { i} }^p \right)  ^{1/p}$ 
bounded by a small constant 
$  \epsilon  $ for some chosen $p$. In the case of image classification the $ l _ { \infty } $ norm 
$\norm{\boldsymbol{\delta}}_{\infty }= \max _ { i } \abs{ \delta  _ { i} }  $ is usually selected, corresponding to perturbations whose pixels are each 
individually bounded in magnitude by $  \epsilon  $. 
Such an attack is said to have ``strength'' $\epsilon$. The $ l _ { \infty } $ norm is the metric we consider in this work. \\ 
\\ \\
\noindent
We generate adversarial perturbations by the method of projected gradient descent (PGD)~\cite{pgd}, which amounts to
attempting to solve the optimisation problem of Equation~\ref{eq:delta_adv} iteratively through a slight 
modification of standard gradient descent. Specifically, starting uniformly randomly in the space of allowable perturbations
$\Delta$, we construct $\boldsymbol{\delta}_{\mathrm{adv}}$ via the procedure
\begin{equation}
  \boldsymbol{\delta}_0' \sim \mathrm{Unif}(\Delta), \ \boldsymbol{\delta}_{t+1}' = \mathcal{P}_{\Delta}(\boldsymbol{\delta}_t' - \alpha \nabla_{\boldsymbol{\delta}} \log(p_{\boldsymbol{\theta}^*}\left(y|\boldsymbol{x} + \boldsymbol{\delta}_t'\right)))
  \label{eq:pgd}
\end{equation}
for a given number of iterations and a step size $\alpha$. 
After each gradient step the perturbation is projected back into $\Delta$ by the projector $\mathcal{P}_{\Delta}$.
Here we take three steps, i.e. $\boldsymbol{\delta}_{\mathrm{adv}}=\boldsymbol{\delta}_{3}'$, and set $\alpha=\epsilon/3$,
where $\epsilon$ is the strength of the attack.
It is a particularly notable result of adversarial machine learning that standard image-classifying convolutional neural networks are vulnerable to being fooled by small perturbations which a human would have no trouble recognising as not meaningfully changing the image~\cite{szegedy2013intriguing} (see Figure~\ref{fig:adv_ims}(a,b). A further surprising result is the \textit{transferability} of adversarial examples -- while the optimisation procedure of Equation~\ref{eq:pgd} involves maximising the loss function of a specific network,  adversarial examples constructed in this way tend to also fool independent networks, even if their architectures differ. We see this  phenomenon in Figure~\ref{fig:adv_ims}(c,d), where adversarial examples constructed with respect to a convolutional neural network also succeed in deceiving a ResNet~\cite{he2016deep} model. The QVCs, however, display a striking robustness to the classically generated adversarial images, as a result of their learning different representations of the features of the images~\cite{west2023benchmarking}. \\

\noindent
The significant reductions in circuit depth attainable by our approximation techniques without large decreases in classification accuracy is possible as a result of the remarkable robustness of the QML models to (non-adversarial) perturbations (Figure~\ref{fig:1}(b) and Figure~\ref{fig:adv_ims}). Further examples of images corresponding to approximately prepared states from the FMNIST dataset, are shown in Supplementary Figure~\ref{fig:noisy_ims}. Despite the visual quality of the images appearing extremely poor, the models are able to discover and exploit subtle properties of the states which are correlated with the class labels of the corresponding images, enabling  them to perform classification with reasonable accuracy (see Figure~\ref{fig:adv_ims}(c,d). Ironically, this inhuman approach to classification of detecting complicated, high frequency correlations instead of analysing the large-scale semantic features of images has been blamed for the existence of adversarial attacks in the first place~\cite{ilyas2019adversarial}. The classification-informing features of the noisy images employed by the QVCs will not, however, be generally the same as those used by a network trained on the exact images, as a result of the noisy preparation procedure failing to recreate their subtle details. This suggests that QVCs trained and evaluated on noisy images may offer increased robustness to adversarial attacks transferred from a network that uses the exact images, as the noisy QVCs will be unable to make use of the features employed by the exact models, which are precisely the features targeted in an adversarial attack. Similar ideas have been proposed in the classical machine learning literature~\cite{cohen2019certified,li2019certified,lecuyer2019certified}. In our case this noise induced resilience combines with the fact that the QVCs naturally learn different classes of features than classical neural networks anyway due to the classical intractability of generic quantum computations~\cite{west2023benchmarking} to produce models that display considerable robustness to classically generated adversarial attacks (see Figure~\ref{fig:adv_ims}). Additionally, we find that this resilience is present when facing attacks conducted on a QVC, with the   approximate models partially resisting the transferred attacks (see Supplementary Figure~\ref{fig:qaml_results}).
\\ \\
\noindent
\large{\textbf{Experimental Demonstration}}
\normalsize
\\ \\
\noindent
Despite the drastic reductions in state preparation circuit depths afforded by the approximation techniques, evaluating our QML models on the MNIST and FMNIST datasets on real hardware remains  out of reach, in part due to the long variational component, which after the simplification of the state preparation subroutine comes to dominate the total CNOT count. 
We can nonetheless explicitly see the progress made towards running such models on a physical quantum computer by conducting noisy simulations with varying degrees of noise, and tracking the performance of the various models. Figure~\ref{fig:shapes}(a) shows the results of noisy simulations of our QML models with a simple noise model which assumes ideal single-qubit gates and two-qubit gates afflicted by depolarisation noise of variable strength $p$. Although this model overlooks the complicated effects present in real systems~\cite{white2020demonstration}, modelling two-qubit gates as failing with probability $p$ is a useful first order approximation. We find that our approximate methods are capable of maintaining their accuracy under noise levels one to two orders of magnitude greater than an exact model generated using Qiskit's standard state preparation technique. 
\\ \\
\noindent
Although this is an encouraging result, it is also valuable to benchmark our techniques beyond a simulation environment, in the presence of real device conditions. 
For example, Ref.~\cite{ren2022experimental} has recently shown that adversarial examples can fool QML models running on real hardware.
In order to facilitate testing on the limited quantum devices available today we introduce a simple two-class dataset which can be tackled by the QVCs with only a few variational layers. This dataset, which we refer to as the  ``Shapes'' dataset, consists of 8$\cp$8 images of squares and crosses (see Figure~\ref{fig:shapes}(b) for a few examples). The images in this dataset can be encoded by means of Equation~\ref{eq:amp_enc} into the state of a five qubit system. We again find that our state preparation methods require significantly fewer CNOT gates than that produced by Qiskit's generic transpilation algorithm (Figure~\ref{fig:shapes}(c)). Notably, these reductions are sufficient to enable us to successfully classify the Shapes dataset on real quantum hardware, in the presence of the associated noise. Results collected on the 27 qubit IBM quantum machine \textit{ibm\_algiers} (Figure~\ref{fig:shapes}(d)) are shown in (Figure~\ref{fig:shapes}(e)), with the approximate methods leading to high classification accuracy and the generic state preparation resulting in circuits deep enough to be overwhelmed by noise. On adversarial examples generated by an $\epsilon=0.1$ attack on ConvNet and transferred to the other listed networks, we observe the quantum models running on \textit{ibm\_algiers} and utilising approximate state preparation are able to resist the attack more successfully than the classical networks. \\ \\
\noindent
\large{\textbf{Summary and Outlook}}
\normalsize
\\ \\
\noindent
The field of quantum machine learning is rapidly progressing and multiple proof-of-concept experimental demonstrations have already been reported in the recent literature~\cite{ren2022experimental, Pannpj2023, Xiangyu2023,PhysRevApplied.16.024051}. Going forward, the implementation of QML on complex datasets relevant for real world applications will require steady breakthroughs on several fronts, including simplification of the complexity of classical data encoding circuits to reduce hardware overheads (such as performed in our work), designing novel QML architectures capable of learning with  superior accuracies, and the implementation of QML models with error mitigation and/or correction protocols. Our work has addressed a key challenge by simplifying arbitrary state preparation which is an expensive but unavoidable component of many QML algorithms, and can be a serious threat to the prospect of asymptotic advantages over classical methods. We have argued that the ability of ML models to cope with surprisingly large amounts of (non-adversarial) noise may be exploited to considerably reduce the complexity of the required state preparation, by simply settling for low fidelity approximations to the target states, and letting the model learn to deal with the ensuing noise. In this fashion we have been able to reduce the gate count for complicated nine qubit states by two orders of magnitude compared to an exact state preparation, while nearly preserving accuracy. Although in some cases the resulting images seem to a human to bear little resemblance to their targets, the QML models are able to learn to classify them correctly. This significant reduction in gate count drastically reduces the hardware resource requirements for the implementation of fault-tolerant QML based on surface code error correction schemes which typically scale very poorly with the circuit size~\cite{west2023towards}.
Furthermore, we have found that the resulting models display an increased resilience to adversarial attacks, due to their decreased reliance on subtle, easily exploitable features of the data.  
\\ \\
\noindent
Despite significant progress on the state preparation front reported here, there still remain open theoretical questions regarding the scalability of QML models and their resistance to dequantisation~\cite{cotler2021revisiting}. For example, the optimal architectures for QML models remain unclear, and will possibly be problem dependent~\cite{schatzki2022theoretical}, in contrast to the universal architectures to which classical neural networks have been converging~\cite{sutton2019bitter,dosovitskiy2020image}. Nonetheless, the ideas explored in this work, combined with the proof-of-concept demonstration on IBM hardware, suggest that there are considerable benefits to sacrificing input state fidelity both for drastically shorter circuit depths on current noisy hardware, and for a natural robustness to adversarial tampering. In the immediate future  several commercially available road-maps indicate that quantum devices will aggressively scale up, with the possibility of thousands of qubits becoming available within the next five to ten years, which when coupled with the anticipation of error rates tracking down below the levels where they can be overcome by sophisticated error mitigation and correction techniques should allow increasingly capable QML demonstrations. The reduction by orders of magnitude in the encoding circuit resource requirements achieved in our work while largely retaining accuracy and even improving robustness is an important milestone along the way. 
\\ \\
\normalsize
\noindent
\textbf{Acknowledgements:} M.W., A.N., F.M.C. and J.H. acknowledge the support of Australian Government Research Training Program Scholarships. The work was supported by funding through the Australian Army Quantum Technology Challenge, and the access to IBM Quantum Devices was provided by the IBM Quantum Hub at the University of Melbourne. 
M.S. was supported by Australian Research Council Discovery Project DP210102831. Computational resources were provided by the National Computing Infrastructure (NCI) and the Pawsey Supercomputing Research Center 
through the National Computational Merit Allocation Scheme (NCMAS).
This research was supported by The University of Melbourne’s Research Computing Services and the Petascale Campus Initiative.
\\ \\
\noindent
\textbf{Data and code availability:}
The data that support the findings of this study are available within the article. The access to source code can be provided upon reasonable request to the corresponding author. 
\\ \\
\noindent
\textbf{Competing financial interests:} The authors declare no competing financial or non-financial interests.

\def\bibsection{\subsection*{\refname}}

\bibliographystyle{naturemag}
\bibliography{./refs}

\clearpage
\newpage

\onecolumngrid

\noindent
\Large{\textbf{Supplementary Information for}}
\noindent
\center {\Large{\textbf{``Drastic Circuit Depth Reductions with Preserved Adversarial Robustness \\ \phantom{................}by Approximate Encoding for Quantum Machine Learning''} }} 
\RaggedRight
\normalsize

\renewcommand{\thefigure}{\textbf{S\arabic{figure}}}
\renewcommand{\theHfigure}{S\arabic{figure}}
\renewcommand{\figurename}{\textbf{Supplementary Fig.}}

\setcounter{figure}{0}

\noindent

\begin{figure}[h!]
\begin{tikzpicture}
  \node at (0,0) {\includegraphics[width=0.85\textwidth]{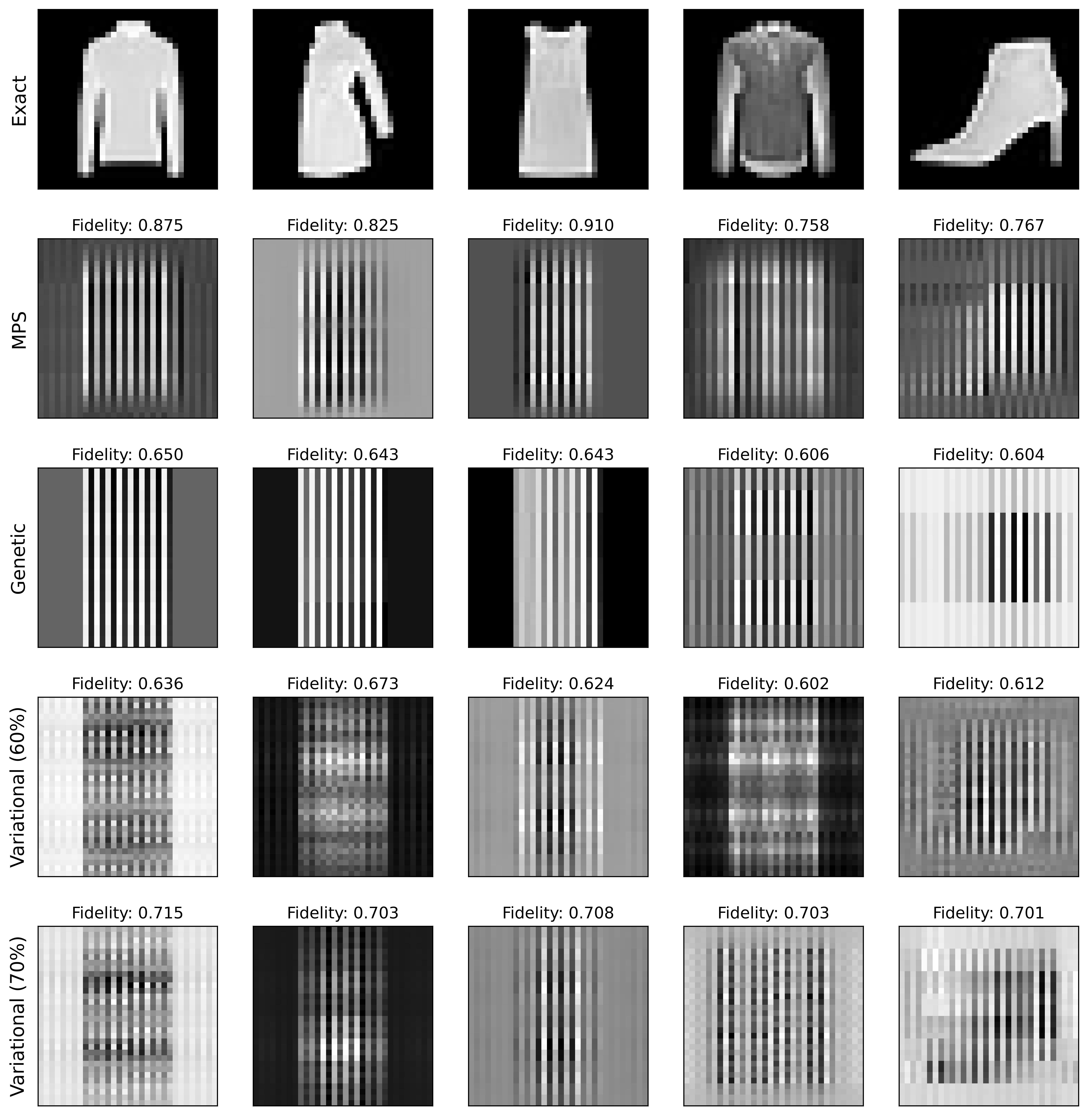}};
\end{tikzpicture}
 \caption{\textbf{Approximately prepared states.} Examples images from the FMNIST dataset~\cite{xiao2017fashion} and images reconstructed from the approximate states prepared by the various methods considered. We find that the visual resemblance of the approximate images is quite poor by human standards, although the machine learning frameworks are still able to carry out the classification with relatively minor reductions in accuracy. The prominent vertical stripes are an artefact of the encoding method (see Equation~\ref{eq:amp_enc}). The effect of alternate  encoding strategies is shown in Supplementary Figure~\ref{fig:alt_encoding}.
 }
 \label{fig:noisy_ims}
\end{figure}

\begin{figure}[h!]
\begin{tikzpicture}
  \node at (0,0) {\includegraphics[width=0.9\textwidth]{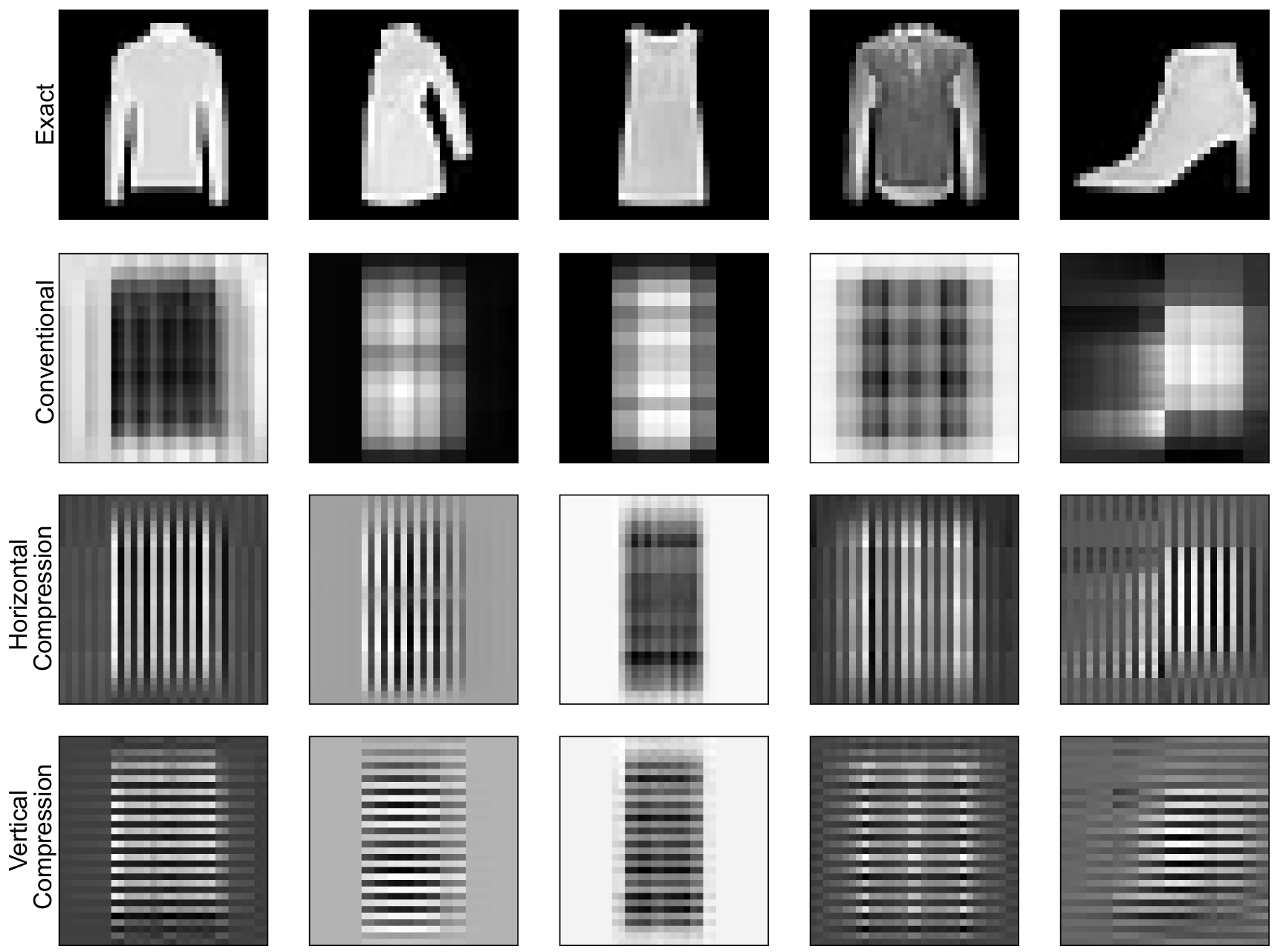}};
\end{tikzpicture}
 \caption{\textbf{Alternate encoding methods.} Our encoding strategy (Equation~\ref{eq:amp_enc}) relies on encoding (horizontally) neighbouring pixels into the real and imaginary components of a given amplitude (``horizontal compression''). Alternately, the real and imaginary components of an amplitude could contain the values of a pair of vertically neighbouring pixels, or each amplitude could simply represent a single pixel value (``conventional'' amplitude encoding). By utilising both the real and imaginary degrees of freedom of each amplitude, the compression methods require one less qubit than standard amplitude encoding. We find, however, that they introduce visual artefacts into the reconstructed images of the approximate states across all methods.
 }
 \label{fig:alt_encoding}
\end{figure}

\begin{figure*}[!h]
\includegraphics[width=0.8\textwidth]{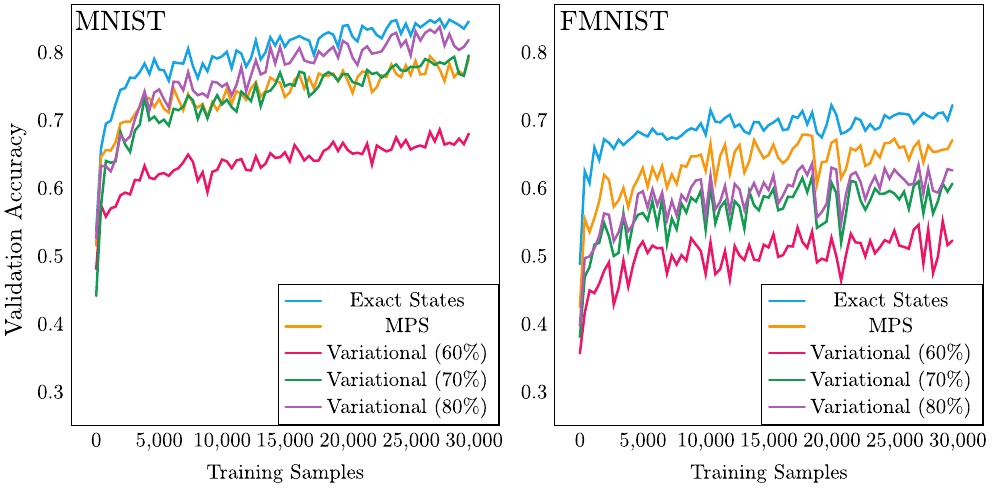}
\caption{The accuracies obtained on a validation set of 500 examples, calculated regularly throughout the training process for QVCs
 trained on images encoded by the various considered methods.}
\label{fig:training_val_accs}
\end{figure*}

\begin{figure*}[!h]
\includegraphics[width=0.7\textwidth]{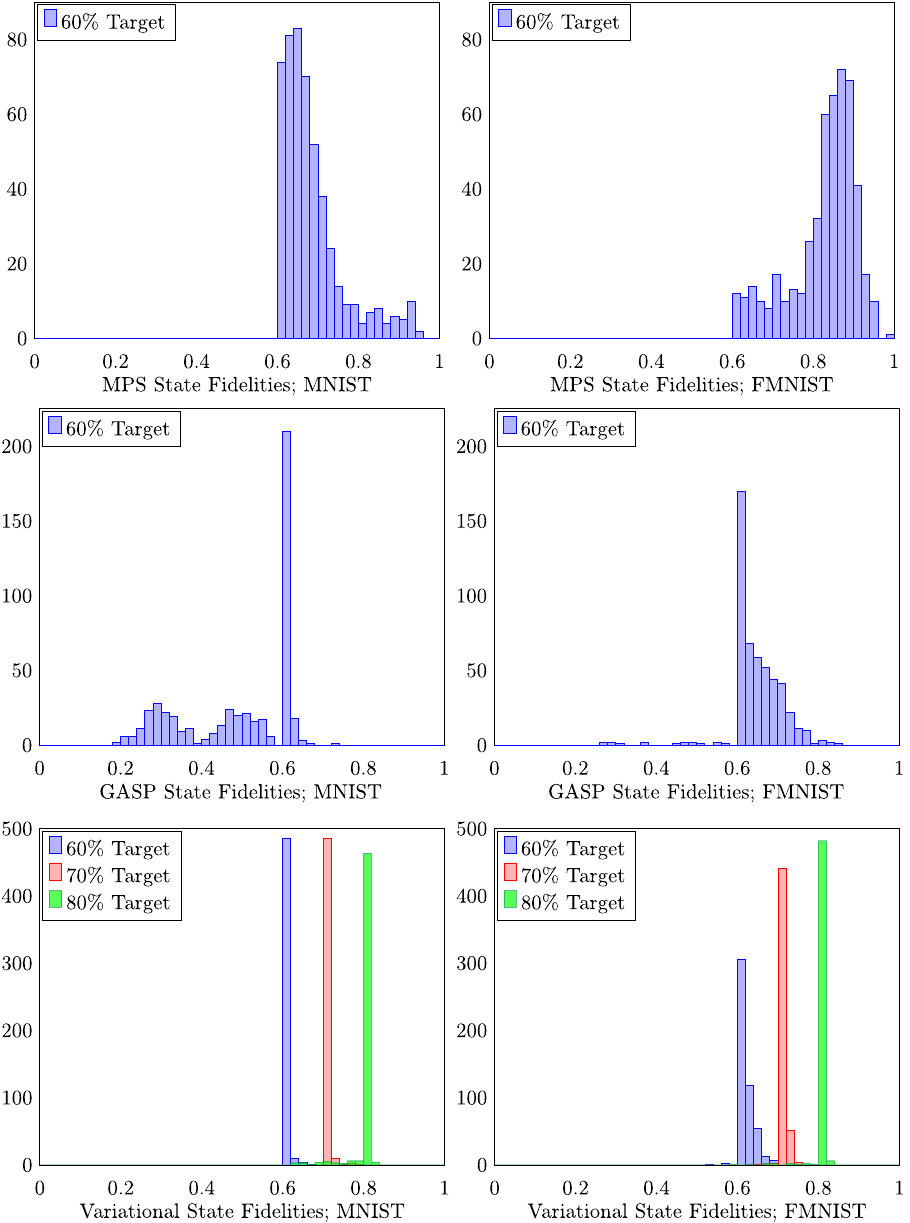}
 \caption{\textbf{Approximate state fidelity.} The fidelities of the approximately prepared states with respect to their targets achieved by the three methods on 500 test images from the MNIST and FMNIST datasets. In the case of the MPS and Variational methods an integer number of layers (see Figure~\ref{fig:1}(d,f)) are always used, which can lead to the target fidelity being overshot. This effect is more pronounced in the MPS case, where a single ``layer'' is more expressive than a layer of the variational circuit. GASP can also overshoot the fidelity target when a mutation leads to a sudden large jump in performance, and occasionally fails to converge, especially in the MNIST case. States corresponding to images from the MNIST dataset are typically harder to produce then states representing the FMNIST dataset. }
\label{fig:val_fidelities}
\end{figure*}

\begin{figure*}
\includegraphics[width=0.9\textwidth]{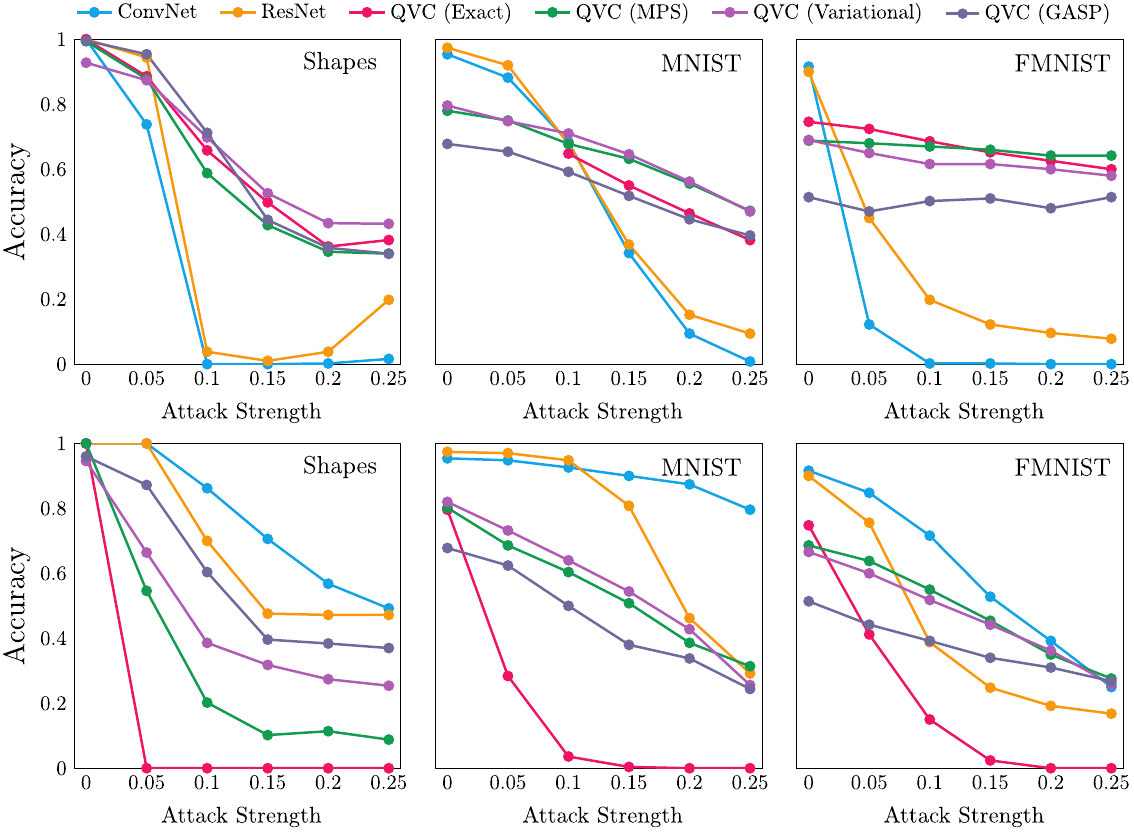}
\caption{\textbf{Adversarial accuracies.} We carry out (PGD) adversarial attacks on a classical convolutional neural network
(ConvNet) and transfer the resulting examples to an implementation of ResNet as well as QVCs employing the 
various considered methods of image encoding. As previously reported~\cite{west2023benchmarking} we find that the QVCs 
are able to resist the transferred adversarial attacks to a far greater extent than ResNet, due to the tendency of 
those classifiers to learn a different set of features to those learned by the classical networks.}
\label{fig:qaml_results}
\end{figure*}

\end{document}